\documentclass[12pt]{iopart}

\pdfoutput=1

\usepackage[LGR,T1]{fontenc}
\usepackage[utf8]{inputenc}
\usepackage[english]{babel}
\usepackage{iopams}
\usepackage{bm}
\usepackage{t1enc}
\usepackage{graphicx}
\usepackage{setspace}
\usepackage{lineno}

\begin{document}

%\begin{frontmatter}

\title{Abnormal subgrain growth in a dislocation-based model of recovery}

\author{Péter Dusán Ispánovity$^1$, István Groma$^2$,\\Wolfgang Hoffelner$^1$ and Maria Samaras$^1$}
%\cortext[cor1]{Corresponding author. Tel.: +41 56 310 4563, fax: +41 56 310 4595. \emph{E-mail address:} peter.ispanovity@psi.ch (P.D. Ispánovity)}
\address{$^1$ Paul Scherrer Institut, CH-5232 Villigen PSI, Switzerland}
\address{$^2$ Department of Materials Physics, Eötvös University, Pázmány Péter stny. 1/a., H-1117 Budapest, Hungary}
\ead{peter.ispanovity@psi.ch}

\begin{abstract}
Simulation of subgrain growth during recovery is carried out using two-dimensional discrete dislocation dynamics on a hexagonal crystal lattice having three symmetric slip planes. To account for elevated temperature (i) dislocation climb was allowed and (ii) a Langevin type thermal noise was added to the force acting on the dislocations. During the simulation, a random ensemble of dislocations develop into subgrains and power-law type growth kinetics are observed. The growth exponent is found to be independent of the climb mobility, but dependent on the temperature introduced by the thermal noise. The in-depth statistical analysis of the subgrain structure shows that the coarsening is abnormal, i.e.\ larger cells grow faster than the small ones, while the average misorientation between the adjacent subgrains remains nearly constant. During the coarsening Holt's relation is found not to be fulfilled, such that the average subgrain size is not proportional to the average dislocation spacing. These findings are consistent with recent high precision experiments on recovery.
\end{abstract}

% \pacs{81.40.Ef, 81.10.Jt, 61.72.Lk, 61.72.Mm}

% \begin{keyword}
% Dislocation dynamics \sep Recovery \sep Abnormal subgrain growth
% \end{keyword}

%\submitto{\MSMSE}
\maketitle

%\end{frontmatter}

%\linenumbers

\section{Introduction}

During the initial stages of annealing of a highly deformed metal, dislocations organize into low angle grain boundaries (LAGBs) that enclose cells (subgrains) with low dislocation content \cite{humphreys}. At elevated temperatures these walls are mobile leading to the gradual coarsening of the subgrains. This phenomenon, known as recovery, has a particularly important role, for instance in the nucleation of recrystallization \cite{humphreys}.

The experimental investigation of recovery is rather difficult, since it is obscured by the early onset of recrystallization, and usually only a relatively small amount of subgrain growth can be observed \cite{sandstrom, varma, furu}. Recently, however, measurements were carried out on specimens of certain orientation and deformation modes, where the recrystallization is suppressed, and much more subgrain growth occurs \cite{ferry_humphreys, huang_humphreys, ferry_burhan}. Due to the development of precise orientation measurement techniques \cite{humphreys_ebsd}, these new experimental results shed new light on the physics of recovery. The most important findings revealed the abnormality of growth, with large subgrains growing faster than small ones \cite{ferry_humphreys, huang_humphreys}, and that the coarsening obeys type 2 kinetics \cite{humphreys, huang_humphreys, ferry_burhan}. The latter means that the time dependence of the average subgrain diameter ($D$) is well-approximated by a power-law
\begin{equation}
D^n-D_0^n = ct,
\end{equation}
with $D_0$ being the cell size at $t=0$ and $c$ being an appropriate constant. The growth exponent $n$ was found to depend strongly on the annealing temperature and on the initial microstructure with values in the range $\sim2-7.5$ \cite{humphreys, huang_humphreys}. Additionally, the size distribution of the grains was observed to be close to lognormal, and its variance was found to increase faster than its average in time \cite{ferry_humphreys, ferry_burhan}.

Several methods have been proposed to model subgrain growth. Molecular dynamics offers the possibility of investigating the dynamical aspects of recovery at the microstructural scale of individual atoms, however, computationally such simulations are too expensive \cite{haslam, haslam2}. For example in the MD simulations performed in Ref.~\cite{haslam} with a volume of $70 \times 70 \times 1.5$ nm$^3$ and containing 25 subgrains of 15 nm diameter, the portion of grain/subgrain growth within the timeframe of 7 ns is quite limited. Such a simulation is at the limit of both time and length scales possible with MD simulations. To overcome these constraints several mesoscopic models have been proposed, including Monte Carlo Potts models (e.g.\ Ref.~\cite{holm, holm2}), phase-field method (e.g.\ Ref.~\cite{ma}) and vertex simulations (e.g.\ Ref.~\cite{humphreys_vertex, weygand_vertex, hayes}) -- all of them relying on input parameters such as grain boundary mobility or energy.

During recovery mainly LAGBs are moving, where the misorientation between adjacent subgrains is less than $\sim 15^\circ$. In this case the dislocation cores do not overlap, thus it is completely justified to study the properties of LAGBs within the framework of dislocation theory (see Ref.~\cite{sutton_balluffi} or Ref.~\cite{gottstein} for reviews). However, only a few results have been published concerning dislocation-based modeling of cell formation and coarsening. They either focused on LAGB mobility measurements (e.g.\ Ref.~\cite{lim}) or on a limited ensemble (e.g.\ Ref.~\cite{argaman}). Recently, a new algorithm was developed by Bakó et al., which, due to a certain multipole method, made it possible to study a much larger number of dislocations than before \cite{bako_2007, bako_prl}. It was shown, that enabling climb leads to cell formation and subsequent subgrain growth with the description of this growth given in terms of power-laws \cite{bako_2007, bako_prl}.

In this paper this two-dimensional (2D) discrete dislocation dynamics (DDD) technique is developed further. As described in section \ref{sec:methods}, the modifications are undertaken in order to increase the computational precision, include thermal noise and use core regularized stress fields to account for dislocation core effects. Due to the improvements, the maximum number of dislocations increases to $N = 200\, 000$ in a single simulation run. A method is also developed in this work to obtain the corresponding orientation field and the reconstructed subgrain structure. In section \ref{sec:results} a detailed analysis of the observed subgrain growth is carried out in terms of experimentally measurable quantities. The comparison of the model predictions with the experimental findings is given in the second part of section \ref{sec:results} before concluding.

\section{Applied methods}
\label{sec:methods}

\subsection{Two-dimensional discrete dislocation dynamics}
\label{sec:2d_ddd}

The simulation of the motion and interaction of individual dislocations is carried out using discrete dislocation dynamics in two dimensions. The main features of the method are summarized as follows. The system consists of parallel edge dislocations which are perpendicular to the plane of the square-shaped simulation area of size $L\times L$. Three possible Burgers vector directions $\pm (\cos(m\pi/3), \sin(m\pi/3))$, $m=0,1,2$ are considered, emulating a hexagonal underlying lattice. Periodic boundary conditions are applied and overdamped dynamics are assumed for the dislocation motion.

The stress field of the dislocations is long-range, and therefore at every time step all the pair interactions have to be taken into account, which results in an $\mathcal{O}(N^2)$ algorithm, where $N$ is the total number of dislocations. This time complexity makes it impossible to study a large number of dislocations ($N>10\,000$). To be able to consider a larger ensemble of dislocations Bakó et al.~used coarse grained stress fields instead of the analytical ones, which allows the direct calculation of the pair interactions to be skipped \cite{bako_2007}. The main idea is to build up a coarse grained discrete Burgers-vector density field $(\alpha_x(l,m), \alpha_y(l,m))$ defined on an $M\times M$ mesh. Let $\alpha_x(l,m)$ and $\alpha_y(l,m)$ denote the $x$ and $y$ coordinate of the net Burgers vector divided by the box area in the box indexed by $(l, m)$. From this field the coarse grained stress field can be determined on the same mesh using a discrete convolutional integral. The approximation here is that the stress field is calculated as if each dislocation was at the centre of its cell, so it corresponds to a first order multipole approximation. For further details of the method, of the used periodic stress fields and of the convolution procedure the reader is referred to \cite{bako_2007}. The most important feature of the method is the reduction of the computational demand which allows the inclusion of a much larger number of dislocations than previously (now up to $200\,000$ compared to a few thousands).

\begin{figure}[!ht]
\begin{center}
\includegraphics[angle=0, width=5cm]{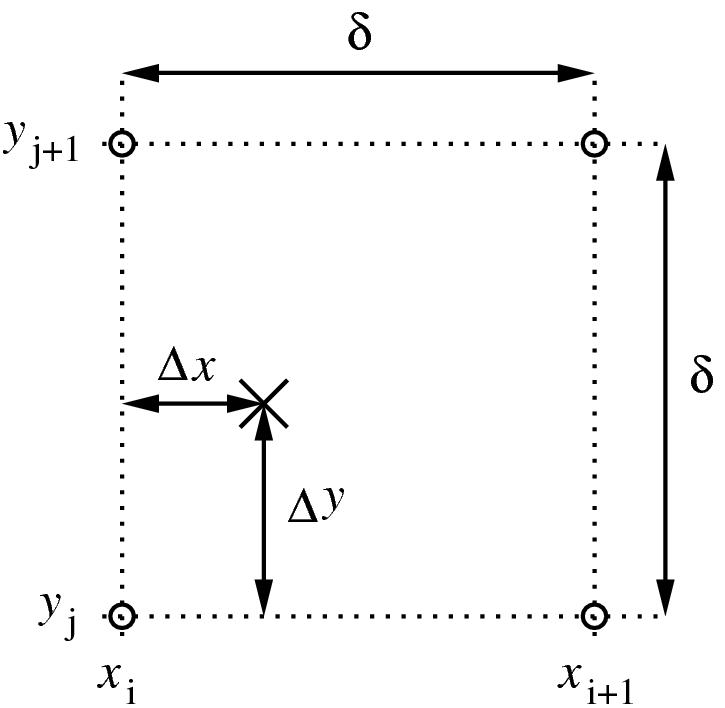}
\caption{\label{fig:interpolation} The schematics of the bilinear interpolation technique. The dislocation position is denoted by the $\times$ symbol, and $\circ$ stands for the grid points of the discrete $M \times M$ mesh.}
\end{center}
\end{figure}

\begin{figure}[!ht]
\begin{center}
\includegraphics[scale=1.2]{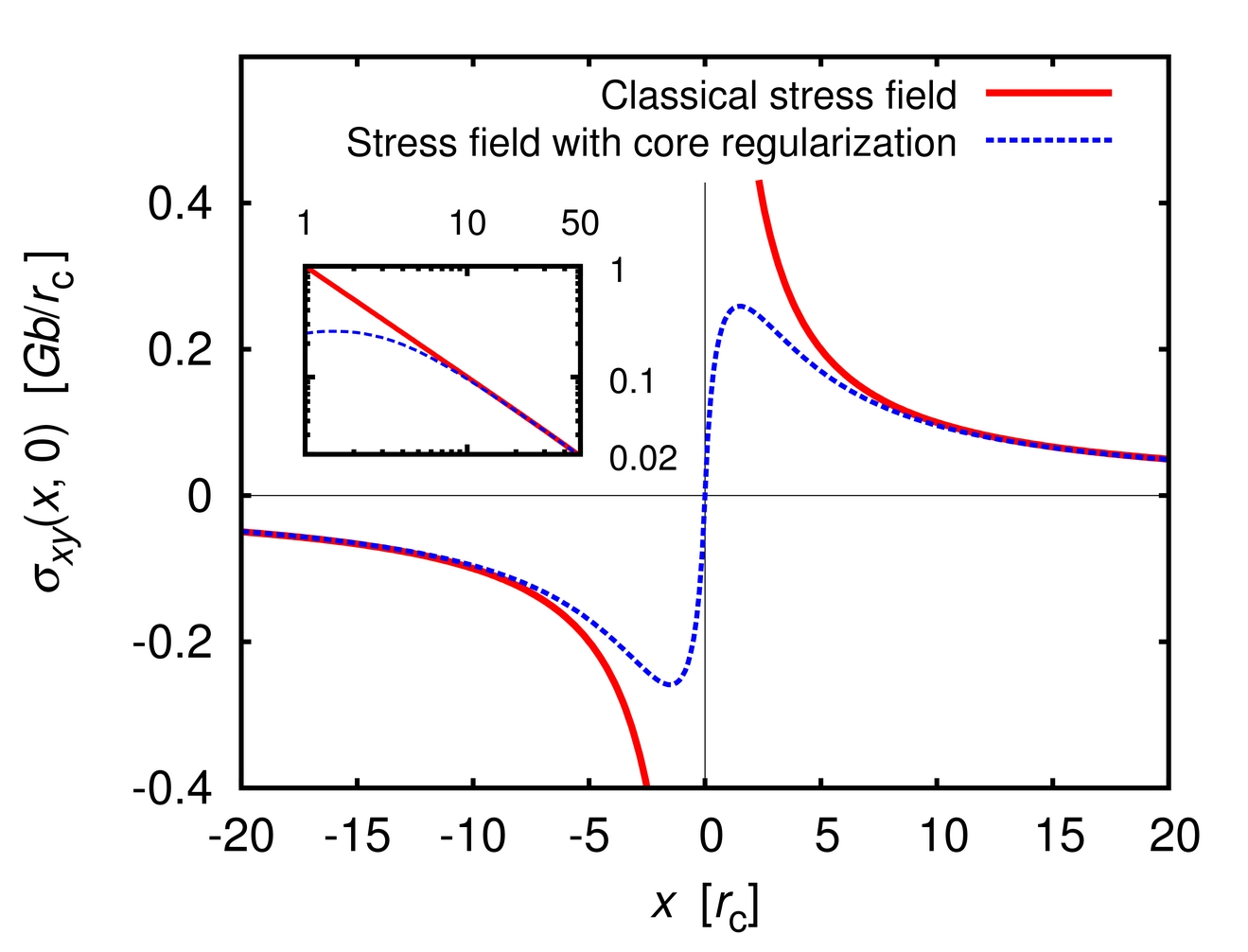}
\caption{\label{fig:tau_xy_sections_actamat} The $xy$ component of the classical and the core regularized stress fields in the glide plane of an edge dislocation. According to the double logarithmic plot of the inset, the difference between the two curves is negligible after a few $r_\mathrm{c}$ values \cite{groma_variation}.}
\end{center}
\end{figure}

In this paper the numerical precision of the $\mathcal{O}(N^2)$ algorithm and the more advantageous time complexity of the coarse grained method is combined. To this end, the following modifications have been included:
\begin{enumerate}
	\item In contrast with the previous method of \cite{bako_2007} where, when determining the $\alpha_i(l,m)$ fields, each dislocation only contributed to the cell which contained it, here the Burgers vector is distributed between the four closest cells weighted with a bilinear approximation term. With the notations of figure~\ref{fig:interpolation}, if the dislocation is positioned between the mesh points $(x_i, y_j)$ and $(x_{i+1}, y_{j+1})$, then the weight for the $(i, j)$ node is $w_{i,j} := (1-\Delta x) (1-\Delta y)/\delta^2$, for the $(i+1, j)$ node $w_{i+1,j} := \Delta x (1-\Delta y)/\delta^2$, etc., where $\delta := 1/M$. As a result, when a dislocation moves from one cell to a neighbour cell, there will be no sudden jumps in the $\alpha_i(l,m)$ field, and therefore, also in the coarse grained stress field.
	\item After performing the convolution, the calculated coarse grained stress field is given on the $M\times M$ discrete mesh ($\tau_\mathrm{cg}(i, j)$). To obtain a smooth (continuous) stress field $\tau_\mathrm{cg}^\mathrm{intp}(x,y)$ we introduce an other bilinear interpolation between the cells as
	\begin{eqnarray}
		\fl \tau_\mathrm{cg}^\mathrm{intp}(x,y) := &\,w_{i,j} \tau_\mathrm{cg}(i, j) + w_{i+1,j} \tau_\mathrm{cg}(i+1, j) + w_{i,j+1} \tau_\mathrm{cg}(i, j+1) \nonumber \\
		&+ w_{i+1,j+1} \tau_\mathrm{cg}(i+1, j+1)
	\label{eqn:interpolation}
	\end{eqnarray}
	with the same weighting factors $w_{i,j}$ as in the previous point. The same is done for the $xx$ and $yy$ component of the stress tensor. The forces acting on dislocations are then computed using the continuous stress field $\tau_\mathrm{cg}^\mathrm{intp}(x,y)$.
	\item To decrease the inaccuracy introduced by the method a length parameter $R_\mathrm{e}$ is introduced. If two dislocations are closer to each other than $R_\mathrm{e}$, their interaction is calculated analytically and not through the coarse grained field. The value of $R_\mathrm{e}$ is chosen to be much larger than the average dislocation spacing.
	\item To account for the fact that close to the dislocation core the generated stress field does not diverge as $1/r$, the core regularized stress fields described in \cite{groma_variation} are implemented. In this case a new parameter, the core radius $r_\mathrm{c}$ is introduced which is of the order of the Burgers vector. In figure~\ref{fig:tau_xy_sections_actamat} the behaviour of the regularized fields is demonstrated by plotting the $xy$ component of the stress field induced by a dislocation in its glide plane.
\end{enumerate}

This modified method is completely continuous, and introduces some approximation only at the interaction of far dislocations. To test the level of the error, the dependency of the results on $R_\mathrm{e}$ has been studied. It was found that if $R_\mathrm{e} \gtrsim 2 \rho^{-0.5}$, then this dependence is negligible (here $\rho$ stands for the total dislocation density and $\rho^{-0.5}$ is the average dislocation spacing). In the rest of this paper this criterion is always fulfilled.

To mimic the effect of temperature, dislocation climb is allowed using the simplest possible mobility rule:
\begin{equation}
	\bm v_\mathrm{g} = M_\mathrm{g} \bm F_\mathrm{g} \quad \mathrm{and} \quad \bm v_\mathrm{c} = M_\mathrm{c} \bm F_\mathrm{c},
\label{eqn:eqn_mot}
\end{equation}
where $\bm v_\mathrm{g}$ and $\bm v_\mathrm{c}$ are the glide and climb velocity of the dislocation, $\bm F_\mathrm{g}$ and $\bm F_\mathrm{c}$ are the glide and climb components of the acting Peach--Köhler force, $M_\mathrm{g}$ is the glide and $M_\mathrm{c}$ is the climb mobility. Since the magnitude of the mobilities can be absorbed into the time scale \cite{csikor_jstat}, the only parameter that affects the dynamics is the ratio $\eta := M_\mathrm{c}/M_\mathrm{g}$ between the two mobilities.

Besides enhancing dislocation climb, temperature has another important effect on dislocation motion: it induces a random movement due to the thermal noise. This should be introduced into the model. Usually it is accounted for by adding a stochastic component to the force acting on the dislocation segments \cite{ronnpagel,raabe, hiratani}. In this work, when the dynamics are overdamped, it is equivalent to adding a stochastic component to the position (Langevin dynamics). This means that after every time step all the dislocations are shifted with $\Delta x_\mathrm{g}$ in the glide, and with $\Delta x_\mathrm{c}$ in the climb directions. The $\Delta x_\mathrm{g,c}$ values are independent, normally distributed random variables with zero mean and, according to the description of Langevin dynamics \cite{risken}, a half width of
\begin{equation}
	\Sigma_\mathrm{g,c}^2 = \frac{2 M_\mathrm{g,c} k_\mathrm{B} T \Delta t}{\Delta l},
\label{eqn:half_width}
\end{equation}
where $T$ is the temperature, $k_\mathrm{B}$ is the Boltzmann factor, $\Delta t$ is the actual time step and $\Delta l$ is the length of the dislocation segment. In this 2D approach it is not possible to define $\Delta l$, therefore, instead of adopting an arbitrary value, in the rest of the paper the material parameters are simply merged into a single dimensionless effective temperature parameter as
\begin{equation}
	T_\mathrm{eff} := \frac{4 \pi (1-\nu) k_\mathrm{B} T}{\mu b^2\Delta l},
\end{equation}
where $\mu$ is the shear modulus and $\nu$ is the Poisson's ratio. Note, that the given half width of (\ref{eqn:half_width}) is the same as in \cite{hiratani}, but slightly different from the results reported in \cite{ronnpagel,raabe}.

An important note has to be made at this point. If the elastic constants and the Burgers vector of for example aluminium at $T = 300$ K are taken with $\Delta l = 10^{-6}$ m as the typical value for the average dislocation spacing, then $T_\mathrm{eff} = 1.3 \times 10^{-4}$ is obtained, which is three orders of magnitude lower than the maximum value used throughout this paper. However, for real 2D crystals with a triangular lattice (the physical system that directly corresponds to the present setup), such as dusty plasmas \cite{quinn_goree, nosenko_zhdanov}, vortex lattices in type II superconducting films \cite{blatter_vortex_review, miguel_zapperi}, colloidal crystals \cite{murray_van_winkle, schall_cohen} and foams \cite{abd_el_kader_earnshaw}, $T_\mathrm{eff}$ can be of the order of 1 due to the difference in material parameters. So, the difference in the $T_\mathrm{eff}$ parameter can be attributed to the fact that a 3D system is modelled in 2D.

The equation of motion (\ref{eqn:eqn_mot}) is solved by a fifth order Runge-Kutta method \cite{numrec}. Annihilation events are introduced if two or three dislocations with zero net Burgers vector are closer to each other than a certain predefined value $d_\mathrm{annih}$.

The presented model is simple and two-dimensional, such that a quantitative agreement with the experiments is beyond its capabilities. However, 2D methods are rather usual for grain/subgrain growth modelling and provide excellent qualitative comparisons with experiments, enabling an insight into the mechanisms at play which cannot be deduced from experiment which will be shown in this paper.

% Nevertheless, the advantage of the model lies in its simplicity, and we have the opportunity to judge which physical phenomenon plays a role in the observed behaviour.

\subsection{Simulation parameters and dimensionless variables}

The simulations are started from random distributions of $N=200\,000$ dislocations. The initial number of dislocations is equal on all three slip planes, and the net Burgers vector is zero. The core radius is set to $r_\mathrm{c}=L/2000$, which is around 20\% of the initial average dislocation spacing, and decreases to around 5\% during a typical simulation. For the coarse graining mesh $M=512$ is chosen. The parameter $\eta$ is varied between $0.02$ and $0.25$, and the effective temperature $T_\mathrm{eff}$ between $0$ and $0.2$. The annihilation distance $d_\mathrm{annih}$ was set to be equal to the core radius $r_\mathrm{c}$.

According to the equation of motion (\ref{eqn:eqn_mot}), the material parameters can be absorbed into the time scale \cite{csikor_jstat}. In contrast to the situation in \cite{csikor_jstat}, here the dislocation density is not constant and a new length scale $r_\mathrm{c}$ has been introduced, so it is advantageous to use $L$ as the normalizing length scale instead of the average dislocation spacing. The dimensionless variables (denoted with prime ($'$)) are, therefore, defined as
\begin{equation}
x' := x/L \quad \mathrm{ and } \quad t' := t M_\mathrm{g} G b^{2}/L^2,
\label{eq:dimless}
\end{equation}
where $b$ is the magnitude of the Burgers vector and $G = \mu / [2 \pi (1-\nu)]$.

\subsection{Characterization of the subgrain structure}
\label{sec:subgrain_structure}

\begin{figure}[!ht]
\begin{center}
\includegraphics[trim = 0 -7 0 0, scale=0.5]{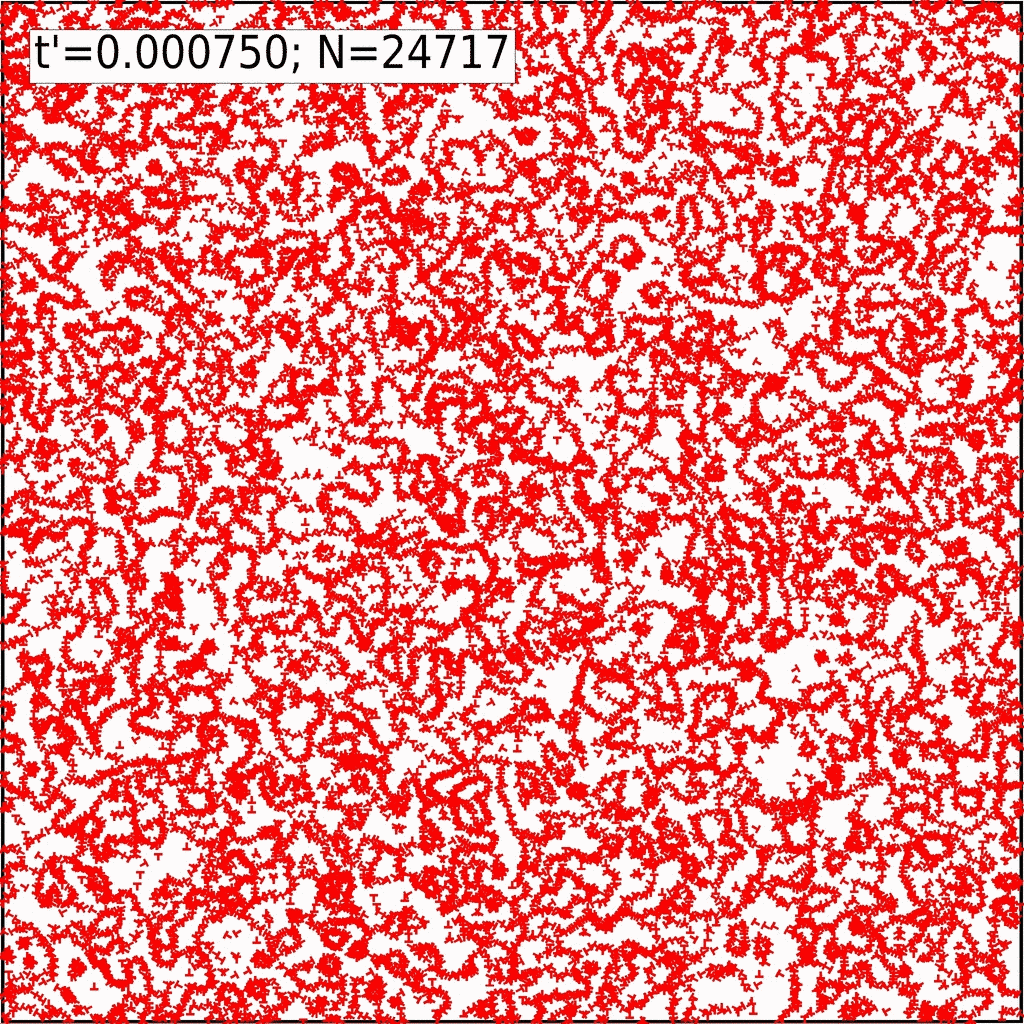}
\includegraphics[trim = 0 -7 0 0, scale=0.5]{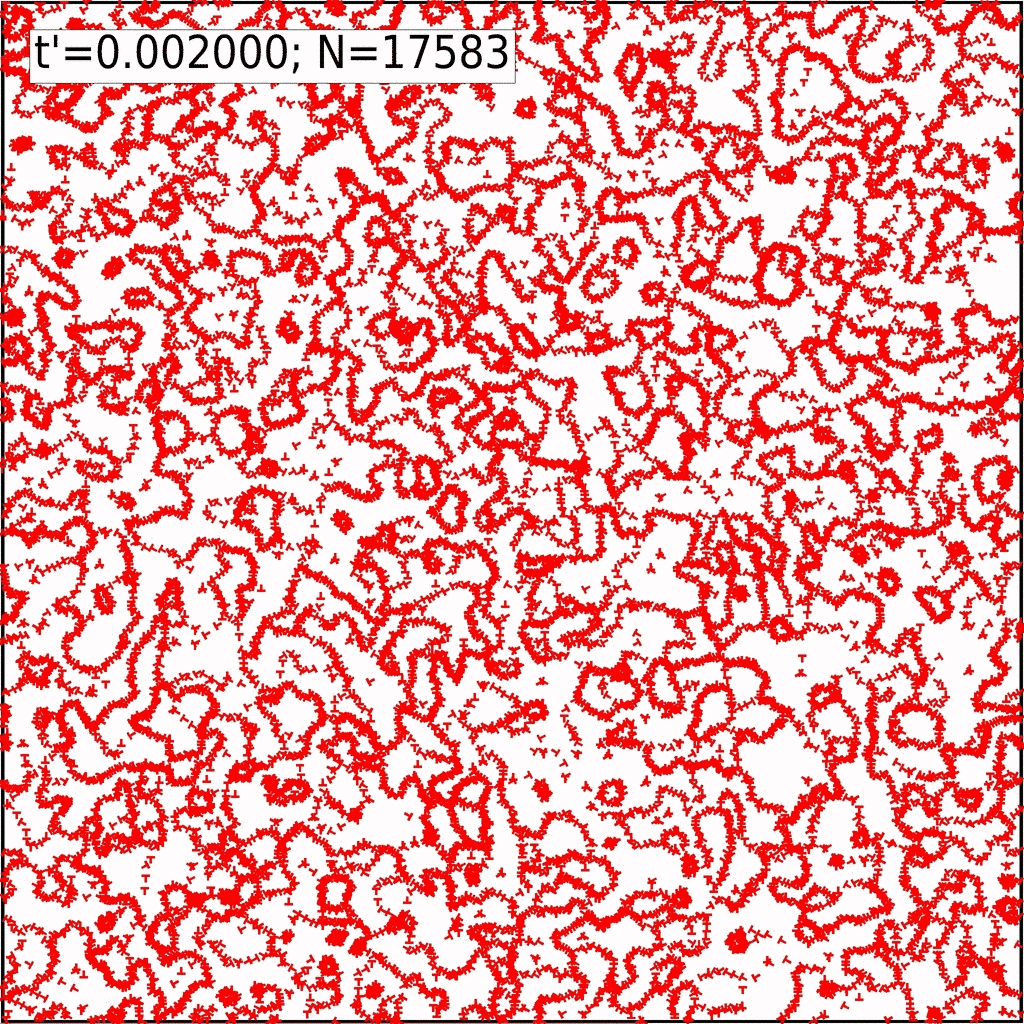}
\includegraphics[trim = 0 -7 0 0, scale=0.5]{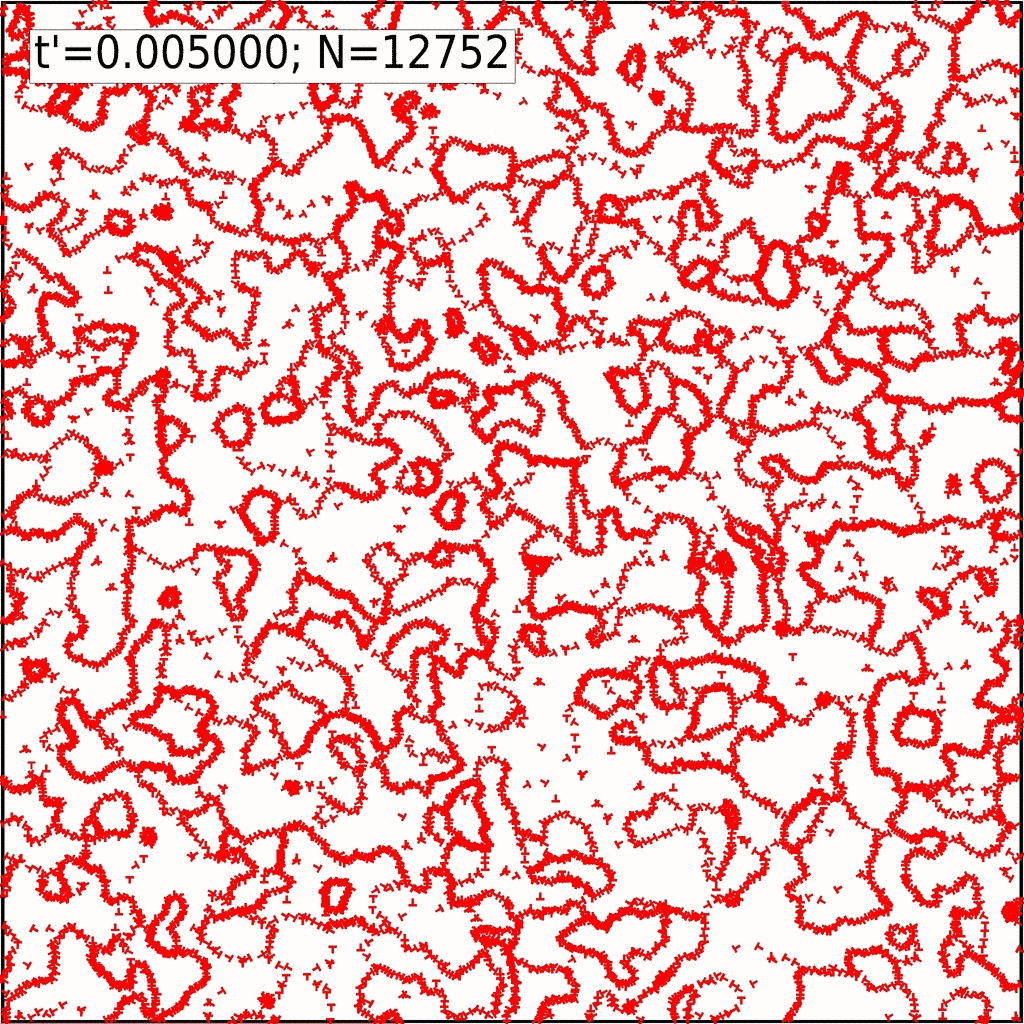}
\includegraphics[trim = 0 -7 0 0, scale=0.5]{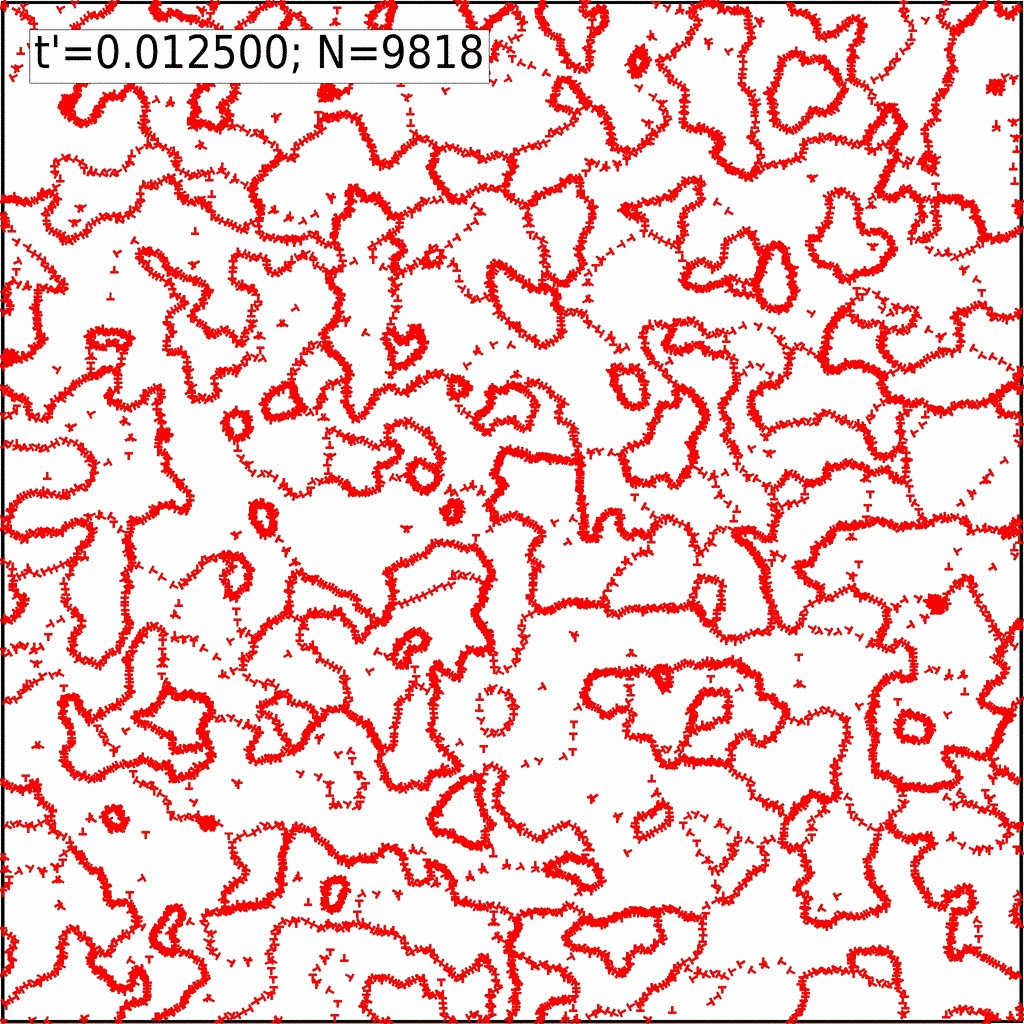}
\includegraphics[trim = 0 -7 0 0, scale=0.5]{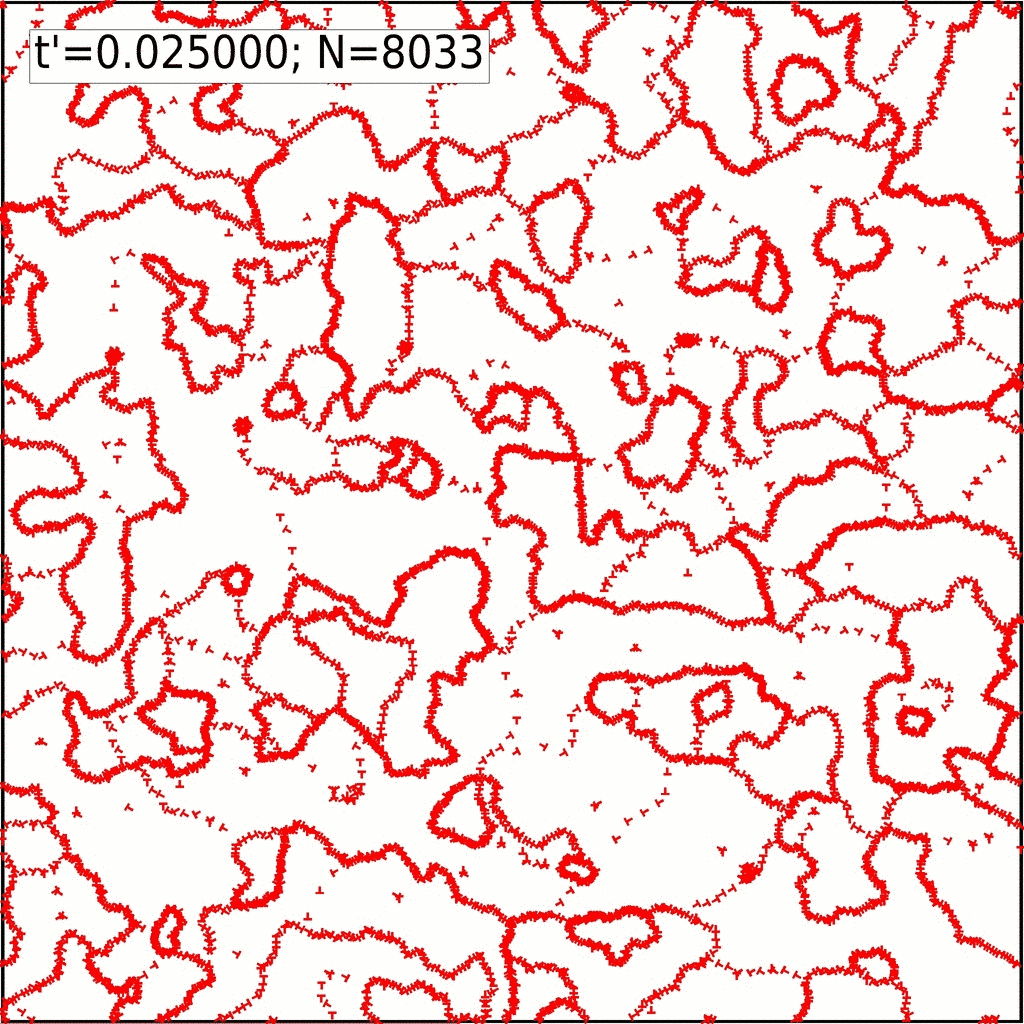}
\includegraphics[trim = 0 -7 0 0, scale=0.5]{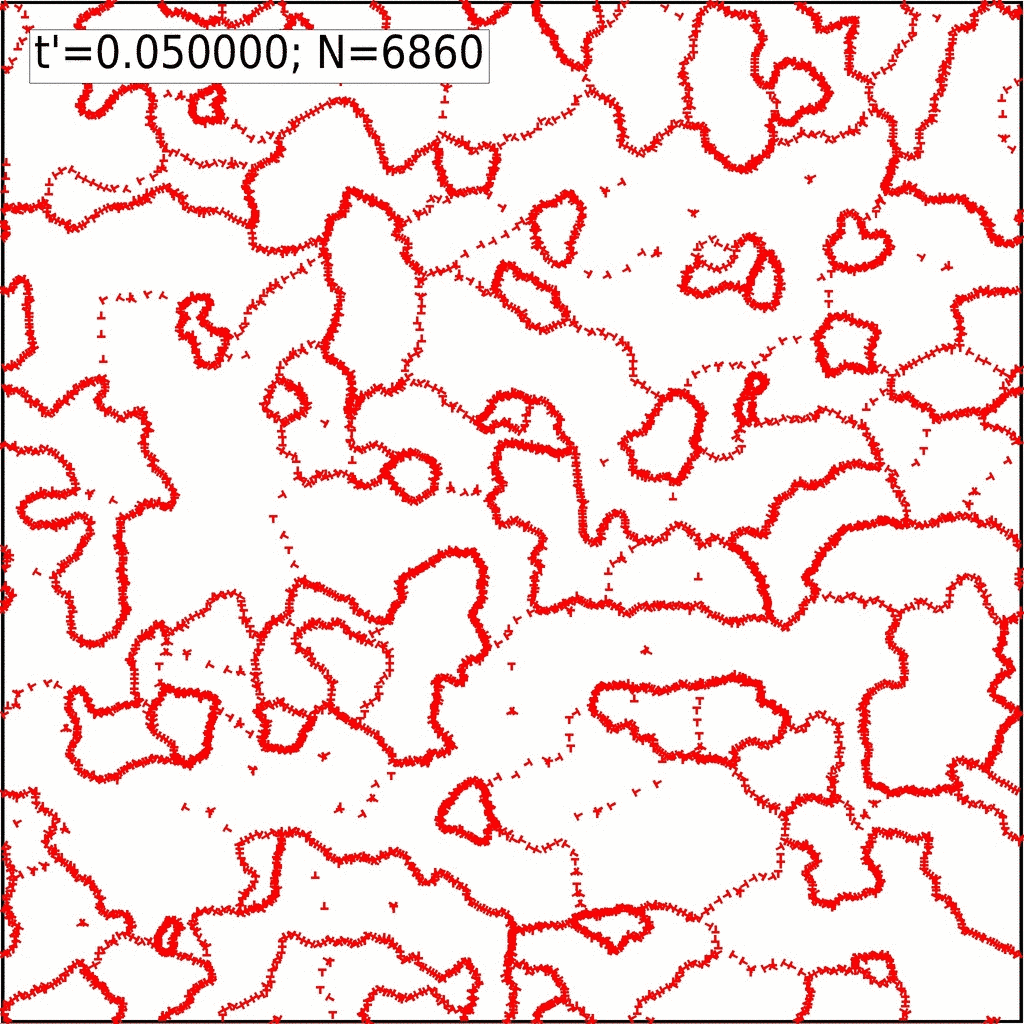}
\caption{\label{fig:sequence} Typical microstructure evolution observed during the simulations. In the images $t'$ denotes the normalized simulation time and $N$ the total number of dislocations (which is equal to $\rho'$ in normalized units).}
\end{center}
\end{figure}

\begin{figure}[!ht]
\begin{center}
\hspace*{0.75cm}
\includegraphics[trim = 0 -7 0 0, scale=0.414]{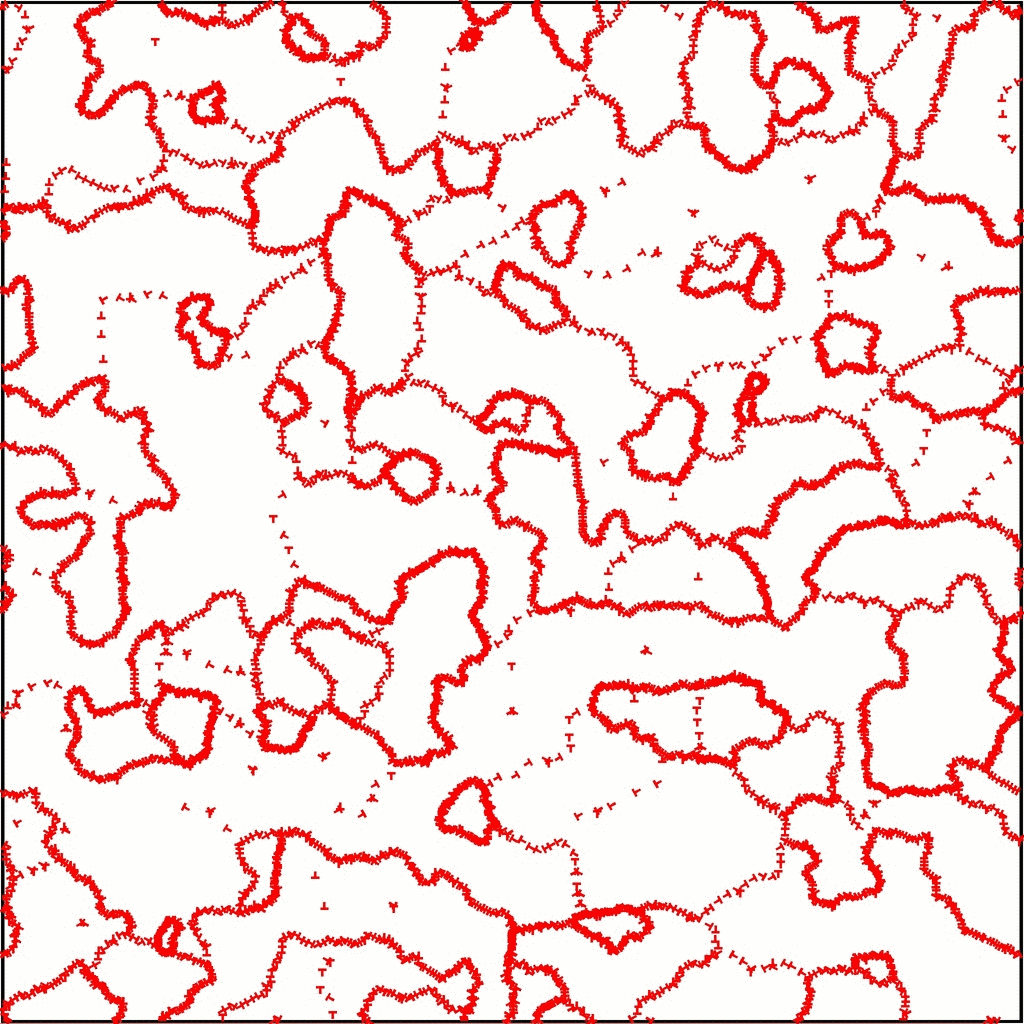}
\begin{picture}(0,0)
% \put(-127,97){\normalsize\sffamily\textbf{a}}
\put(-127,97){\normalsize{(a)}}
\end{picture}
\hspace*{0.5cm}
%\hspace*{1.3cm}
%\includegraphics[trim = -30 0 0 0, scale=0.9]{fig2b_orient.png}
\includegraphics[trim = -30 0 0 0, scale=0.9]{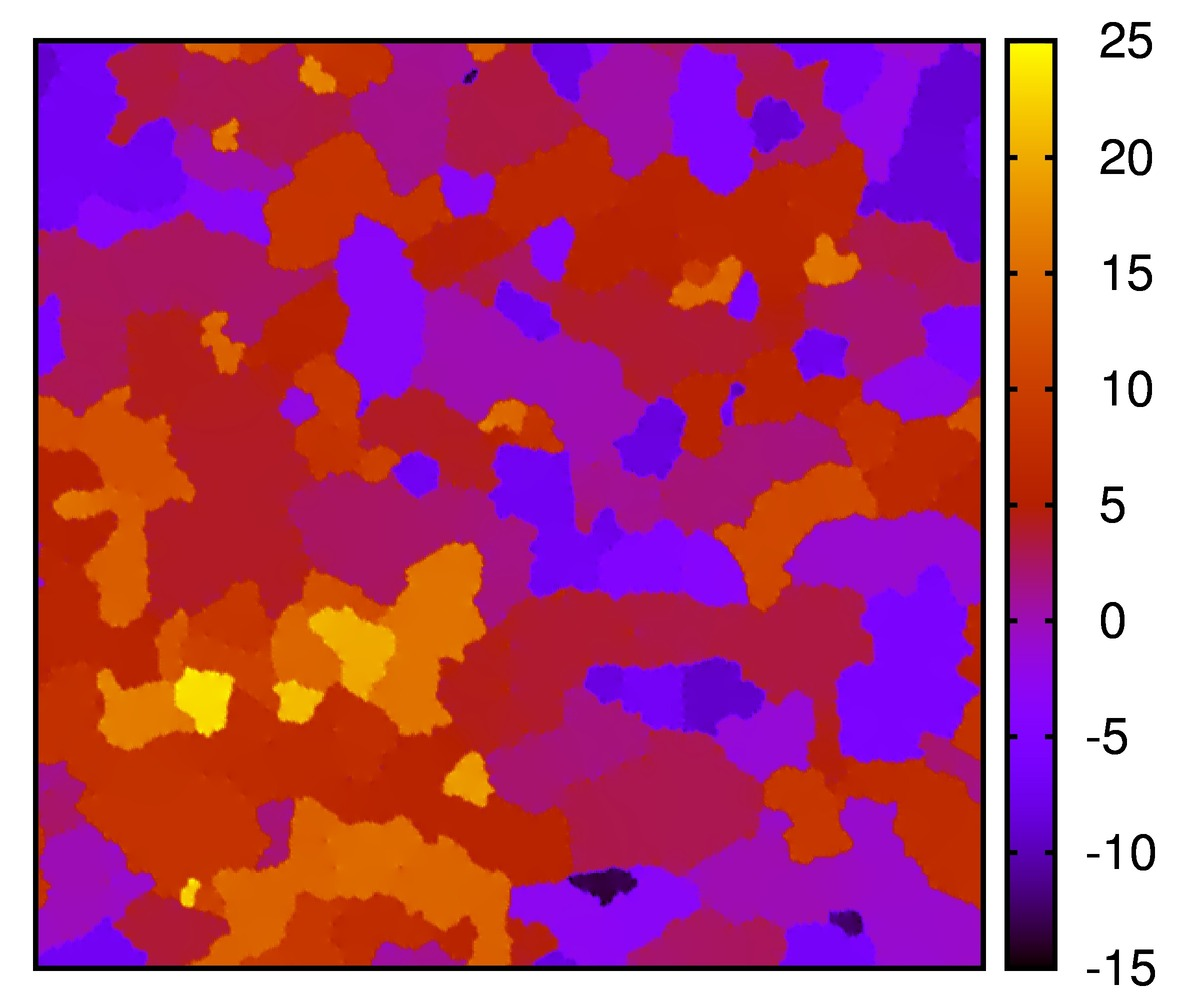}
\begin{picture}(0,0)
% \put(-155,97){\normalsize\sffamily\textbf{b}}
\put(-155,97){\normalsize{(b)}}
\end{picture}
\\
\includegraphics[scale=0.9]{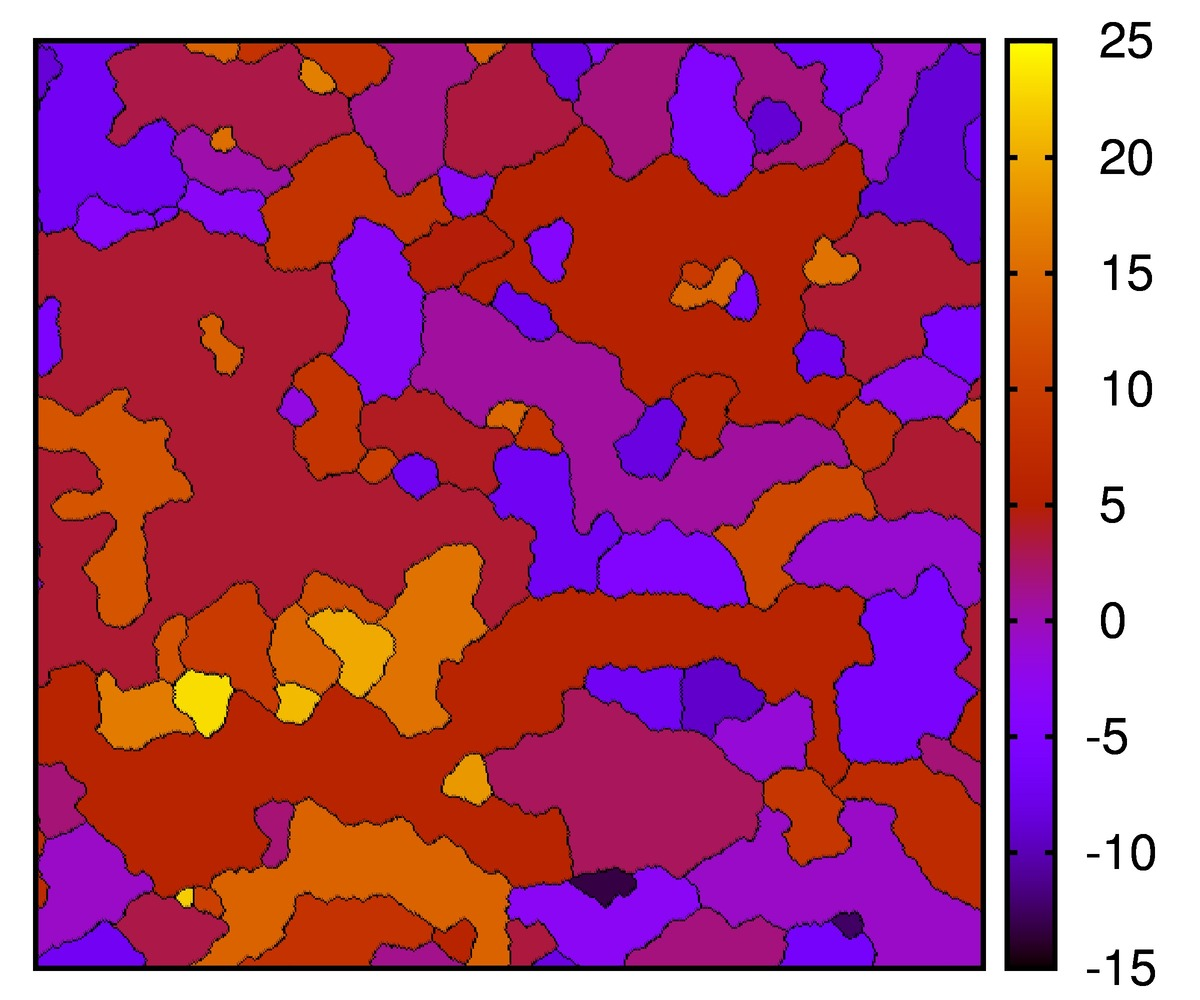}
\begin{picture}(0,0)
% \put(-155,97){\normalsize\sffamily\textbf{c}}
\put(-155,97){\normalsize{(c)}}
\end{picture}
\includegraphics[scale=0.9]{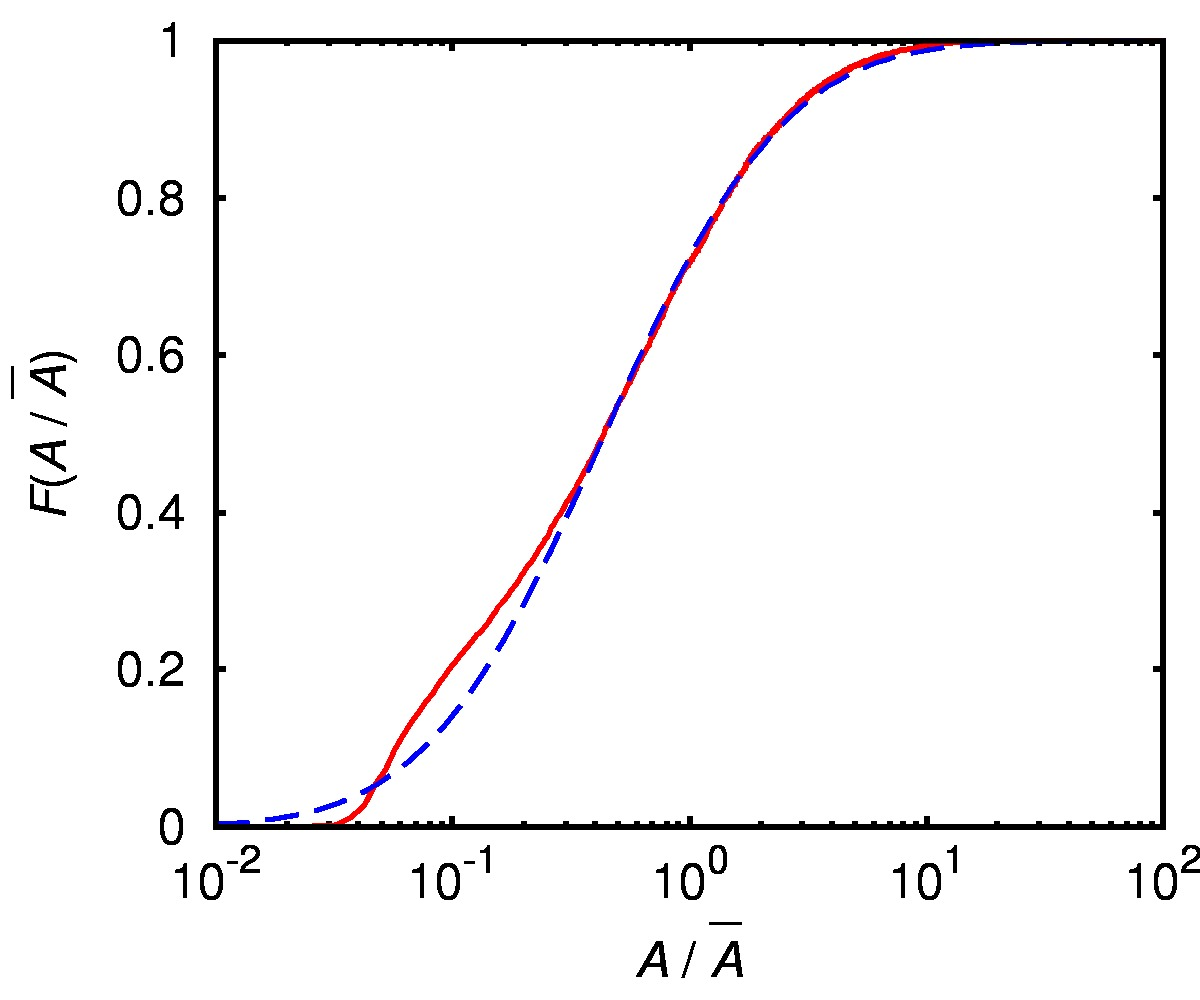}
\begin{picture}(0,0)
% \put(-135,97){\normalsize\sffamily\textbf{d}}
\put(-135,97){\normalsize{(d)}}
\end{picture}
\caption{\label{fig:subgrain_structure} The identification of the subgrain structure. (a) A sample dislocation configuration, (b) the corresponding orientation map (in degrees), (c) the identified subgrains (the colours refer to the average orientation within the subgrains) and (d) the averaged subgrain size distribution. The continuous curve is the measured distribution and the dashed one is the fitted lognormal distribution.}
\end{center}
\end{figure}
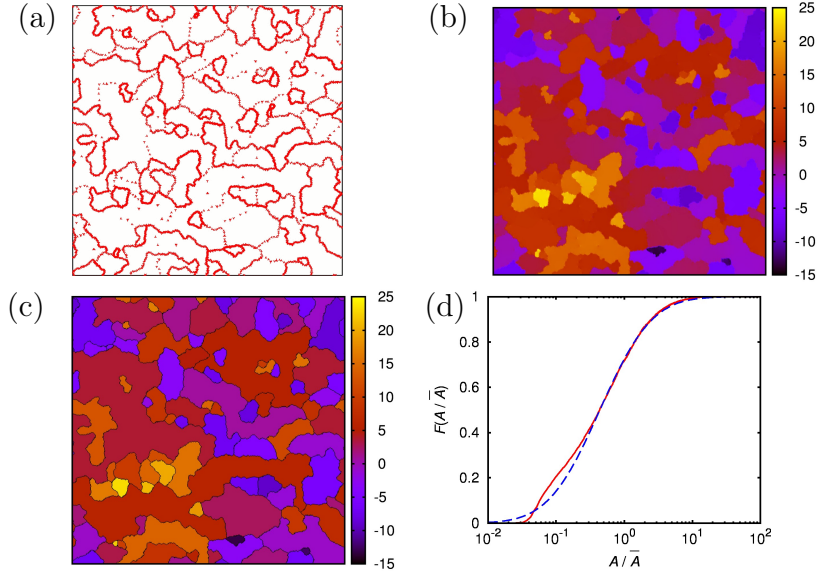

During the simulations subgrains form and then grow, in agreement with earlier simulations \cite{argaman, bako_2007}. A typical sequence of dislocation configurations observed (if dislocation climb is present) is plotted in figure~\ref{fig:sequence}.
Although subgrain formation is apparent, it is not straightforward to define the distinct subgrains from the position data of the individual dislocations. The method applied is now described (and is similar to the one of \cite{argaman}). According to the Kröner-Kosevich continuum theory of dislocations, in the present case, the orientation field $\omega(\bm r)$ is connected to the dislocation density tensor $\hat{\alpha}$ as \cite{kroner_2}
\begin{equation}
	\partial_x \omega = \alpha_x \mathrm{, \ and } \ \  \partial_y \omega = \alpha_y,
\label{eqn:omega_1}
\end{equation}
where the notations $\alpha_x := \alpha_{31}$ and $\alpha_y := \alpha_{32}$ are introduced.\footnote{ In the Kröner-Kosevich continuum theory a quantity called the `dislocation part of the relative rotation' is introduced, which is the non-elastic component of the rotation field, and is denoted by $\bm{\theta}$. The introduced $\omega$ field corresponds to the $z$ component of this field.} These quantities are the continuum versions of $\alpha_{x, y}(l,m)$ defined in section \ref{sec:2d_ddd}. Equations (\ref{eqn:omega_1}) can be transformed into a Poisson equation
\begin{equation}
	\triangle \omega = \partial_x \alpha_x + \partial_y \alpha_y,
\label{eqn:omega_alpha}
\end{equation}
which can be solved efficiently on a discrete grid iteratively by a multigrid method \cite{numrec}. Consequently, periodic boundary conditions are automatically fulfilled. Thus from the dislocation positions the discrete $\alpha_{x,y}(l,m)$ fields can be constructed, and then $\omega(l,m)$ follows from (\ref{eqn:omega_alpha}). This procedure can be performed on an arbitrarily smooth $K\times K$ mesh. Figure~\ref{fig:subgrain_structure}(b) shows the orientation map corresponding to the configuration of figure~\ref{fig:subgrain_structure}(a) for $K=512$. It is important to note that without the introduction of core regularized stress fields it is not possible to define $b$ in a straightforward manner, and therefore, the $\theta \approx b/d$ misorientation cannot be defined. In the present setup, however, one can assume $b := r_\mathrm{c}$, leading to defined $\omega$ values in figure~\ref{fig:subgrain_structure}(b). So, the misorientation values are directly comparable to the experiments.

The subgrain structure is then derived by locating close to constant orientation regions on the $\omega$ field. Figure~\ref{fig:subgrain_structure}(c) shows the structure derived from figure~\ref{fig:subgrain_structure}(b), the colours referring to the average orientation in the subgrain. According to figure~\ref{fig:subgrain_structure}(c), the large subgrains are determined confidently, while there is some arbitrariness in the definition of the small ones. This is the direct reflection of the unavoidable uncertainty of the subgrain definition. After the subgrain structure has been reconstructed the average area $\overline{A}$ and the standard deviation $\delta A$ can be measured. The linear size is then $\overline{D} := \sqrt{\overline{A}}$ and $\delta D := \sqrt{\delta A}$.

The subgrain sizes are of course not uniform, they obey a certain distribution $P(A)$, where $A$ denotes the area of a subgrain. From the reconstructed structure of figure~\ref{fig:subgrain_structure}(c) $P(A)$ is easily obtained. For better numerical accuracy the cumulative distribution function $F(A)$ of $P(A)$ was constructed and then averaging was performed over statistically equivalent realizations of the system. In figure~\ref{fig:subgrain_structure}(d) $F(A)$ is plotted, and one can conclude from the fit that the distribution is close to lognormal (just like in experiments \cite{ferry_humphreys, ferry_burhan}). Note that here and in the rest of the paper, ensemble averaging is always performed over 16 independent simulation runs.

\section{Results}
\label{sec:results}

\subsection{Power-law growth}
\label{sec:power-law}

\begin{figure}[!ht]
\begin{center}
\includegraphics[scale=1.2]{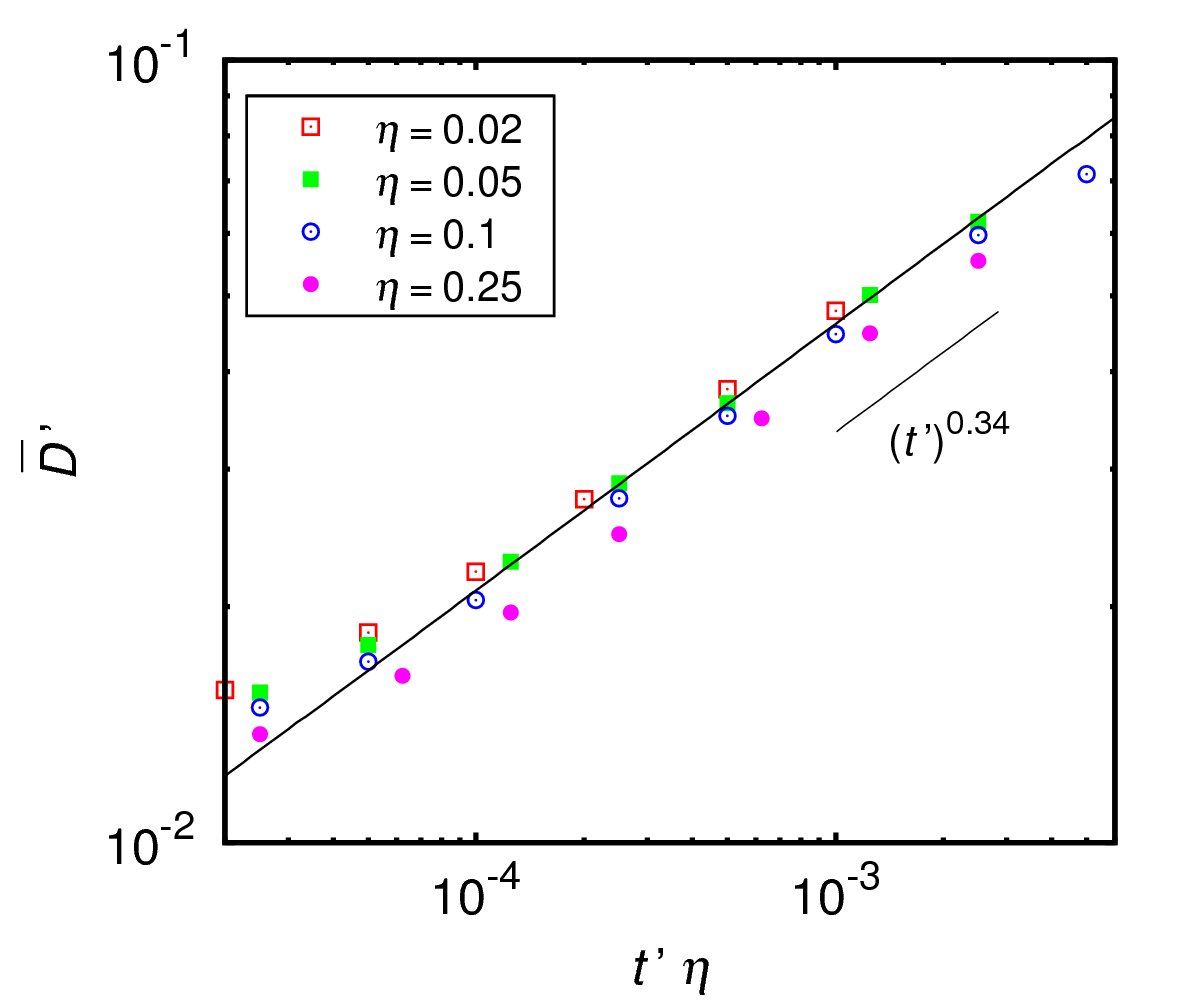}
\caption{\label{fig:davg_eta} The average subgrain size evolution at $T_\mathrm{eff} = 0$ and different climb mobilities. The values were plotted against $t' \eta$.}
\end{center}
\end{figure}

First the effect of the climb mobility is investigated, whereby simulations with different $\eta$ values have been performed at $T_\mathrm{eff} = 0$. As seen in figure~\ref{fig:davg_eta}, the average subgrain size $\overline{D'}$ follows power-law kinetics. Since the smallest identifiable subgrain is limited by the resolution of the mesh on which (\ref{eqn:omega_alpha}) is solved, at small $t'$ values a higher average is obtained than the real value. On the other hand, when $\overline{D'} \approx 0.1$, the size of the largest subgrains is already in the order of the system size $L$, which biases the average in the other direction. Nevertheless, the power-law regime lasts for more than $1.5$ decades, which confirms that the coarsening obeys type 2 kinetics. From the fit $n=2.9\pm 0.2$ is obtained, which is not far from the experimentally observed low-temperature values \cite{humphreys, huang_humphreys}.

The other consequence of figure~\ref{fig:davg_eta} is that the power-law exponent does not depend on $\eta$. More precisely, a slight deviation is only seen when $\eta$ is larger than $0.1$. This is consistent with the results of Hartmaier et al. \cite{hartmaier} who studied the creep properties of thin films. Therefore $\eta=0.1$ is implemented in the rest of the work presented in this paper. Moreover, the curves almost overlap when they are plotted versus $t' \eta$ indicating that the climb rate only modifies the time scale of the simulation but not the nature of the dynamics.

\begin{figure}[!ht]
\begin{center}
\includegraphics[scale=1.2]{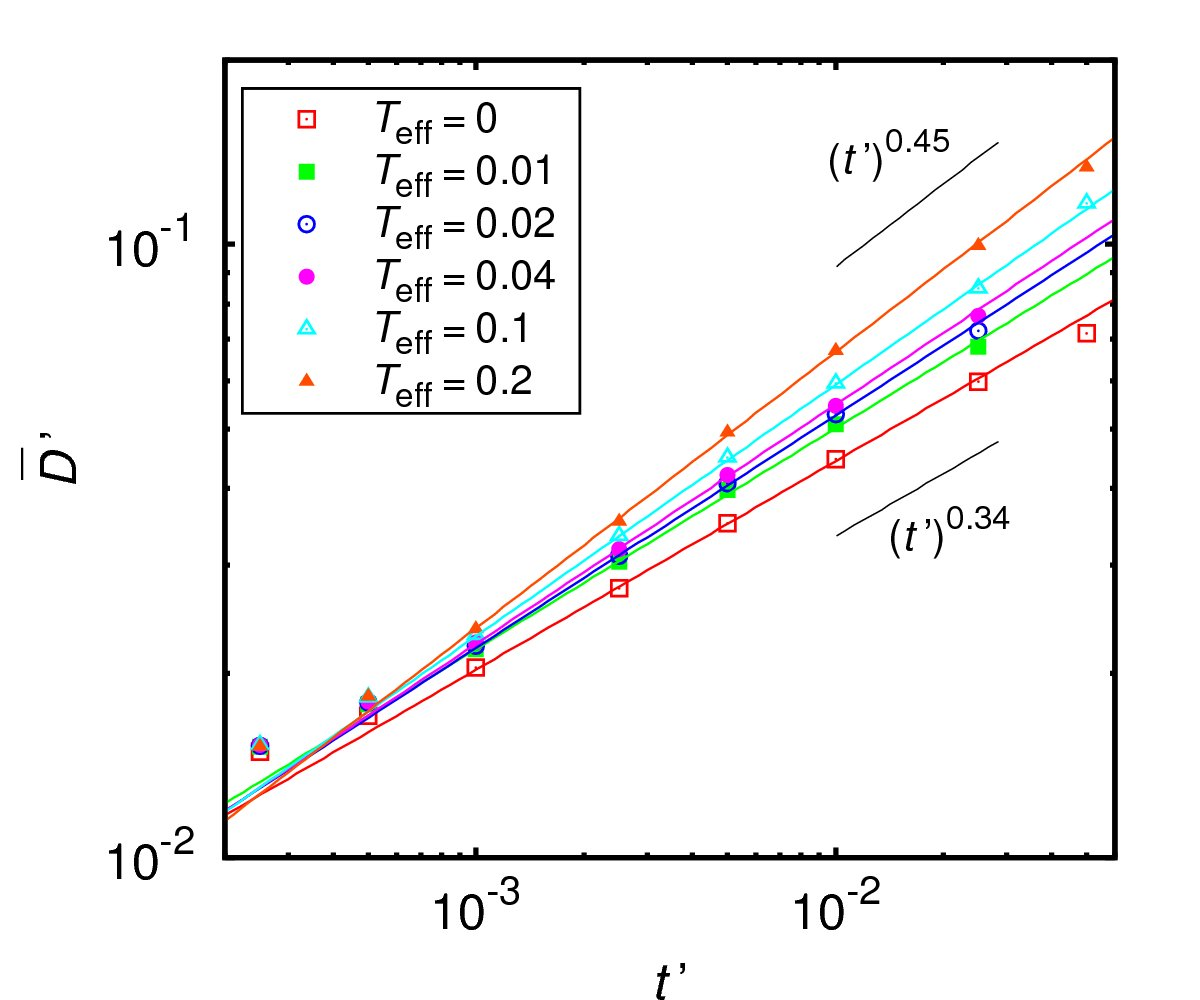}
\caption{\label{fig:davg_temp} The average subgrain size as a function of time at $\eta=0.1$ and different $T_\mathrm{eff}$ values.}
\end{center}
\end{figure}

The constancy of the power-law exponent contradicts experimental findings which report decreasing $n$ with increasing temperature \cite{humphreys, huang_humphreys}. However, the addition of thermal noise to the dislocation motion (as described in section \ref{sec:2d_ddd}) leads to the expected increase of the power-law exponent, that is, to the decrease of $n$. This is demonstrated in figure~\ref{fig:davg_temp}, where it is seen that the exponent of $n=2.9\pm 0.2$ decreases to about $2.2 \pm 0.2$. The numerical values of the exponents are different from the experimental values, which is one of the limitations faced by a 2D model studying a 3D phenomenon. In addition, several mechanisms that hinder boundary motion, such as dislocation-point defect interactions, are not considered. The tendency observed, however, is in good agreement with the experiments \cite{humphreys, huang_humphreys}.

\begin{figure}[!ht]
\begin{center}
\includegraphics[scale=1.2]{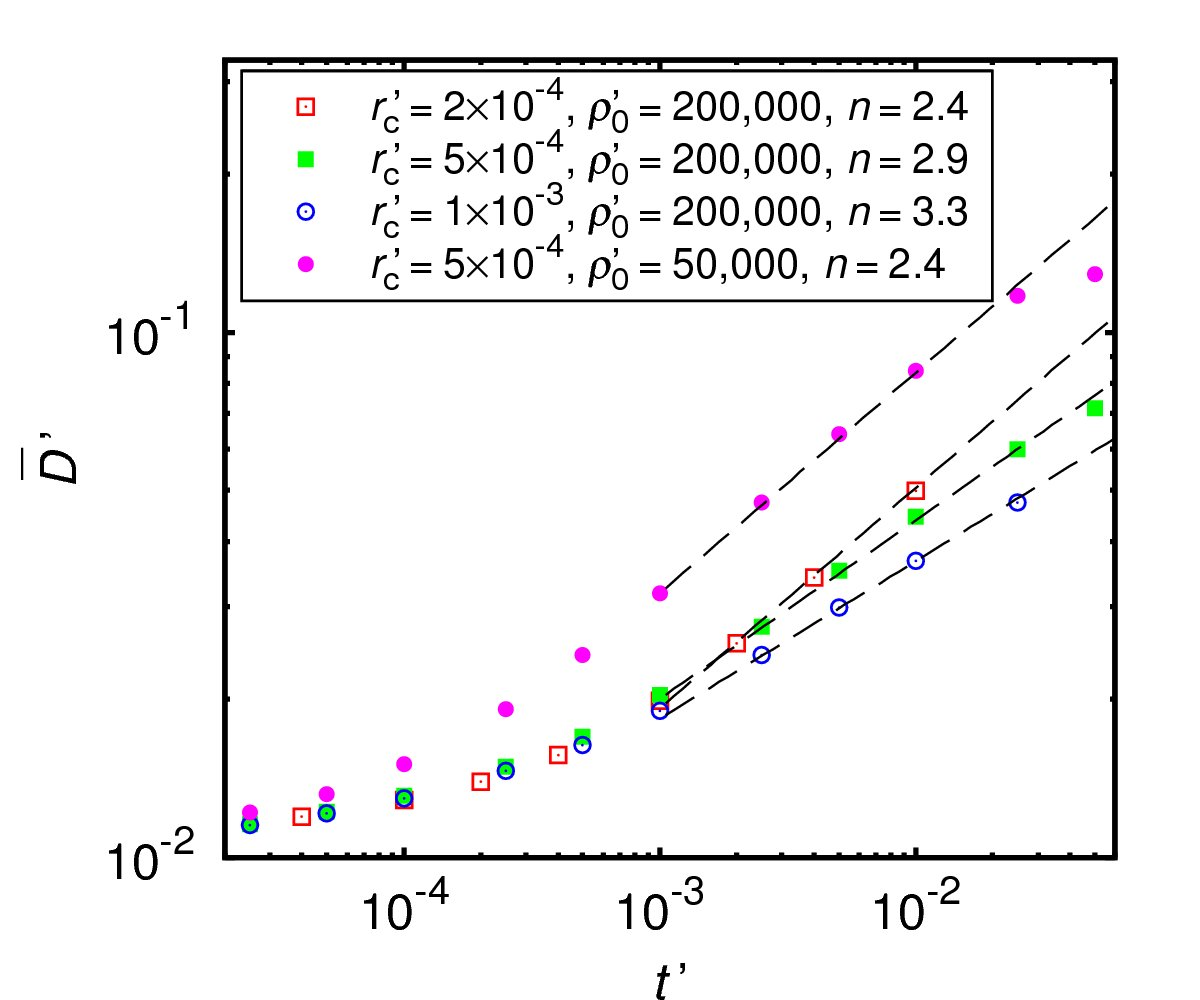}
\caption{\label{fig:davg_rc} The average subgrain size as a function of time at $\eta=0.1$ and different $r_\mathrm{c}$ and $\rho_0'$ values.}
\end{center}
\end{figure}

The exponent was found to depend on the initial number of dislocations $N_0$ as well as on the core radius $r_\mathrm{c}$. As seen in figure~\ref{fig:davg_rc}, $n$ increases as the core radius and the initial density grow and as expected, only the $r_\mathrm{c}/\rho_0^{-0.5}$ ratio, i.e.\ the ratio between the core radius and the average dislocation spacing, is important. A possible reason for this dependence could be that the mobility of a dislocation wall increases with decreasing core radius. However, for a thorough explanation further studies are required.

\begin{figure}[!ht]
\begin{center}
\includegraphics[scale=1.2]{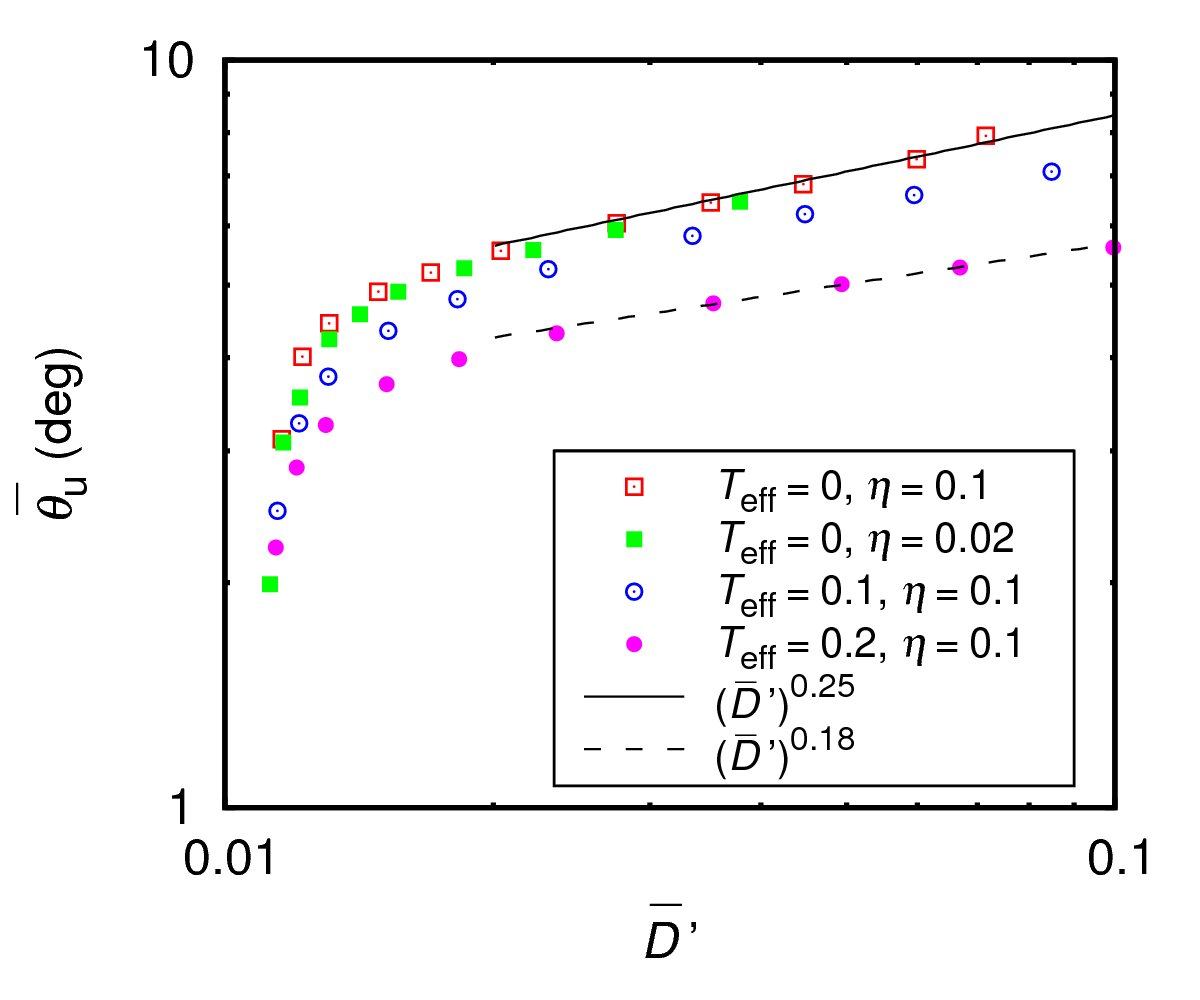}
\caption{\label{fig:davg_misorient} The average misorientation as a function of the average subgrain size.}
\end{center}
\end{figure}

It is also instructive to look at the evolution of the average misorientation between adjacent subgrains. Starting from the reconstructed subgrains of figure~\ref{fig:subgrain_structure}(c) the mean uncorrelated misorientation $\overline{\theta}_\mathrm{u}$ was measured which is by definition the mean misorientation between all possible neighbour subgrain pairs. As seen in figure~\ref{fig:davg_misorient}, a slowly growing average misorientation was found which is non-sensitive to the climb mobility and its growth rate decreases with increasing temperature. In experiments $\overline{\theta}_\mathrm{u}$ was found to be nearly constant \cite{huang_humphreys}. In this work the growth rate is low, and is even decreasing at high temperatures, thus providing good qualitative agreement.

The discussion now returns to growth exponents. It was found above that the exponents $n$ are slightly below the experimentally obtained values. They are surprisingly high, however, if compared to the predictions of the existing models of subgrain growth kinetics. Models based on dislocation climb predict parabolic subgrain growth, i.e.\ $n=2$ \cite{humphreys, furu, sandstrom_1977_1, sandstrom_1977_2}. As a possible solution for the discrepancy it was suggested that the thermally activated migration of the ledges is the rate controlling mechanism \cite{furu}, or the gradual decrease of the average misorientation is responsible for the increased exponent \cite{huang_humphreys} and not the climb. According to these simulation results it is evident that a simple climb model is able to account for an exponent of $n=3.3$, even with a slightly increasing average misorientation. Thus it is likely that some important feature of subgrain growth is missing from the existing climb-based models, which should, therefore, be revised.

\subsection{Discontinuous subgrain growth}

During discontinuous (or abnormal) subgrain growth large cells grow faster than small ones, which leads to an inhomogeneous subgrain distribution during annealing. This type of growth was confirmed by recent experiments \cite{ferry_humphreys, huang_humphreys}. A simple consequence of this behaviour is that the subgrain structure is not self-similar, and the scaled size distribution of subgrains broadens with time.

\begin{figure}[!ht]
\begin{center}
\includegraphics[scale=1.2]{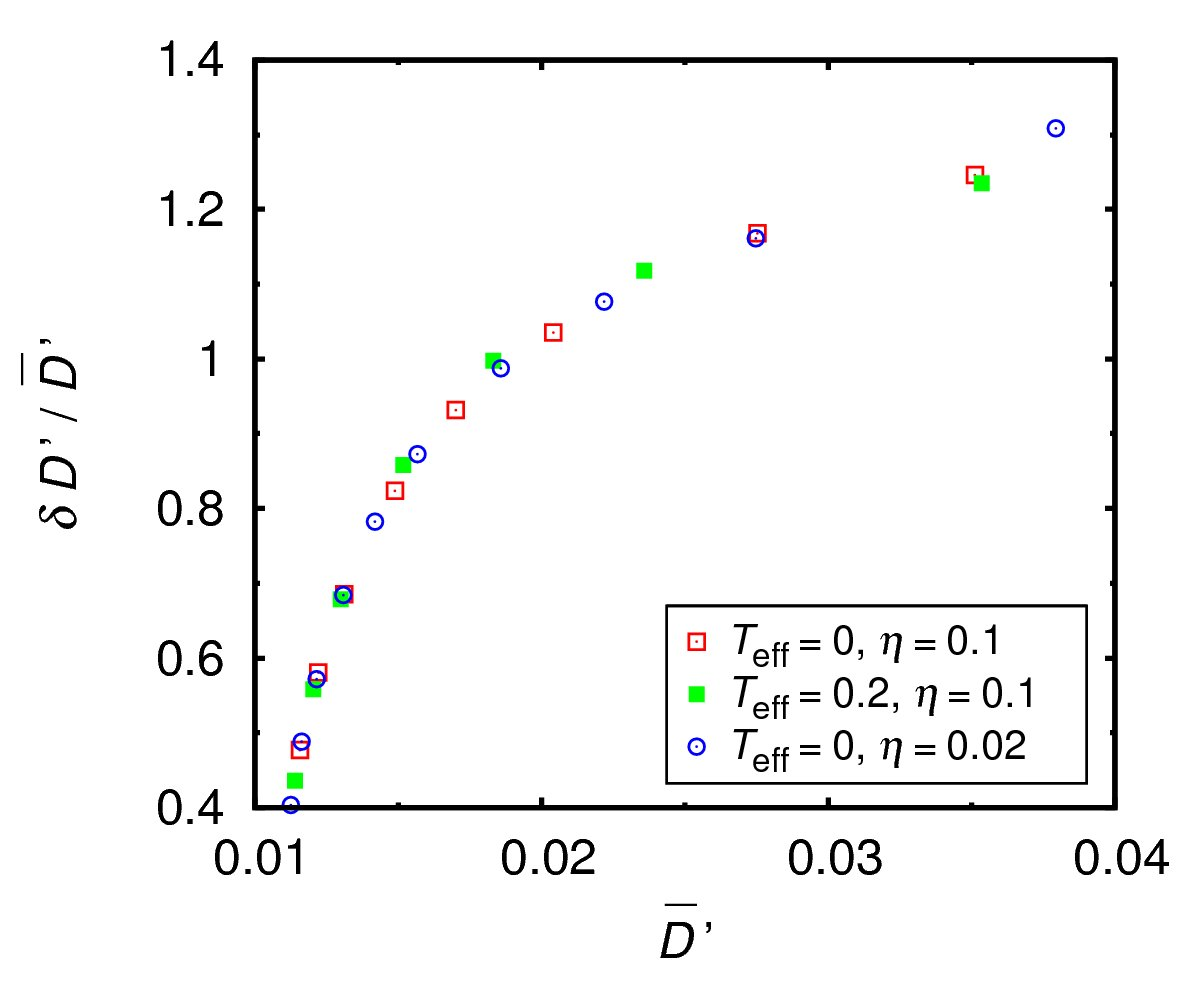}
\caption{\label{fig:davg_sigma_rel} The relative standard deviation of subgrain sizes versus average subgrain size.}
\end{center}
\end{figure}

From the reconstructed subgrain structures the average linear size of the subgrains ($\overline{D}'$) and its standard deviation ($\delta D'$) were measured as described in section \ref{sec:subgrain_structure}. In figure~\ref{fig:davg_sigma_rel} the evolution of the relative scattering $\delta D'/\overline{D'}$ is plotted at various simulation parameters. It is evident that the growth is discontinuous, and that neither the temperature nor the climb rate has any effect on it.

\subsection{Length scales in the microstructure}
\label{sec:length_scales}

It is commonly assumed that in large dislocation systems there is only one length parameter: the average dislocation spacing $\rho^{-0.5}$. This means that dislocation structures are scalable, which in the case of cell sizes is expressed by Holt's relation \cite{holt}
\begin{equation}
	D = K \rho^{-0.5},
\label{eqn:holt}
\end{equation}
where $K$ is a suitable constant. Relation (\ref{eqn:holt}) was proven experimentally for example on deformed iron \cite{holt}, on crept LiF \cite{reppich} or more recently on doped GaAs \cite{rudolph}. It is surprising, therefore, that Holt's relation (\ref{eqn:holt}) contradicts the observed abnormal nature of subgrain growth. Namely, if there is only one length parameter (the system is scalable) only self-similar, i.e.\ normal growth, is possible. But in the present case to describe the size distribution at least two length parameters are needed (its mean and its deviation) since the observed relative scattering of cell sizes increases, and therefore the growth rate cannot be normal.

\begin{figure}[!ht]
\begin{center}
\includegraphics[scale=1.2]{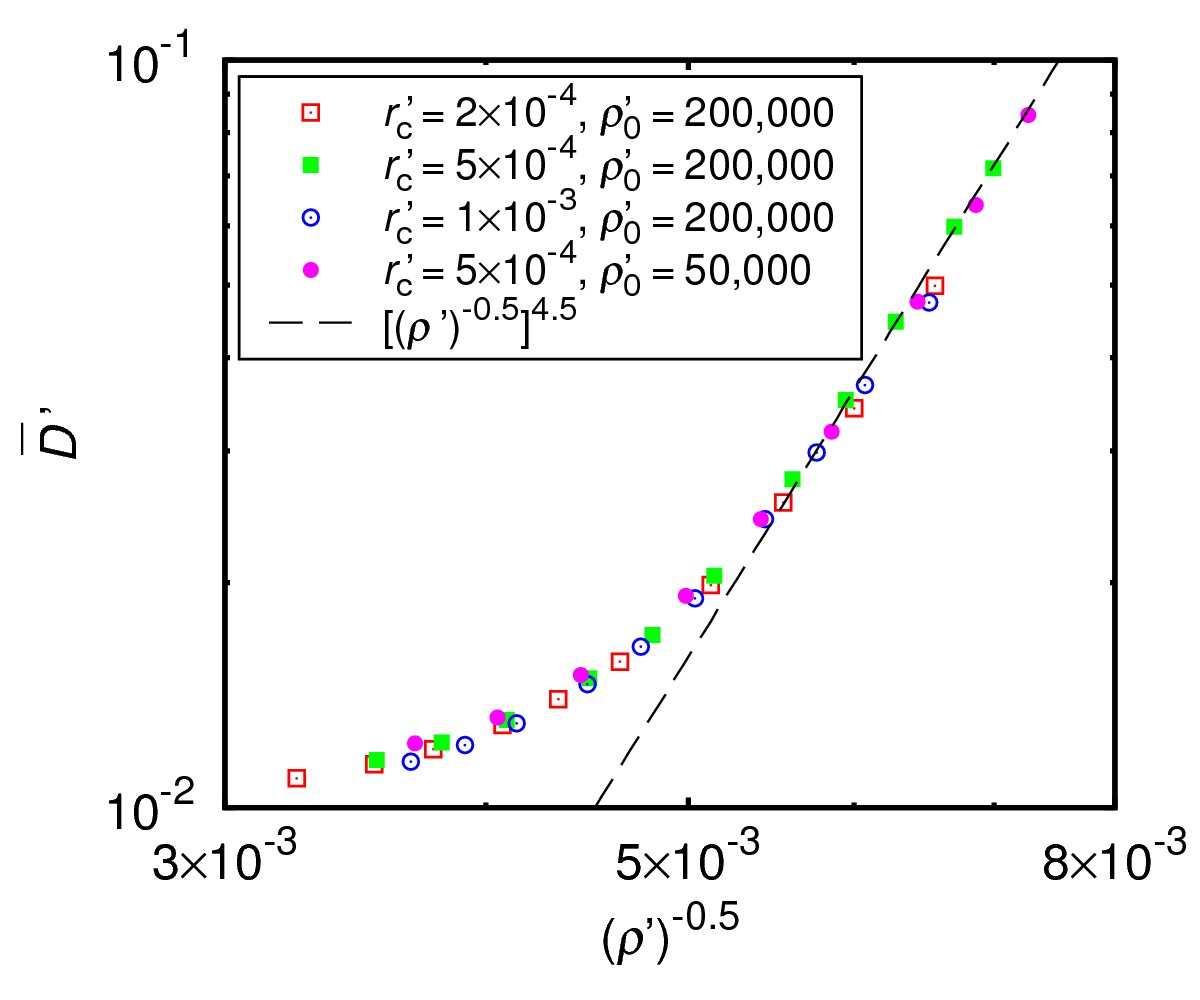}
\caption{\label{fig:ddisloc_davg} The average cell size ($\overline{D'}$) versus the average dislocation spacing ($\rho^{-0.5}$). Holt's relation would imply a linear dependence.}
\end{center}
\end{figure}

It was already mentioned that in these simulations together with the core regularized stress fields a new length parameter, the core radius $r_\mathrm{c}$ was introduced. Since Holt's scaling argument only assumes a $1/r$ type dislocation stress field, this could solve the contradiction. Hence, the simulations were repeated with different core radii, and the average cell diameter $\overline{D}$ dependence on the average dislocation distance $\rho^{-0.5}$ was investigated. Figure \ref{fig:ddisloc_davg} shows that the effect of $r_\mathrm{c}$ is negligible on the $\overline{D} - \rho^{-0.5}$ relation. This is understood as the average distance of dislocations in the LAGBs was found to be at least 3 times larger than the core radius $r_\mathrm{c}$, even at the highest value of $r_\mathrm{c}$. So, it is not the introduction of core regularized fields that breaks Holt's relation.

\begin{figure}[!ht]
\begin{center}
\includegraphics[scale=1.2]{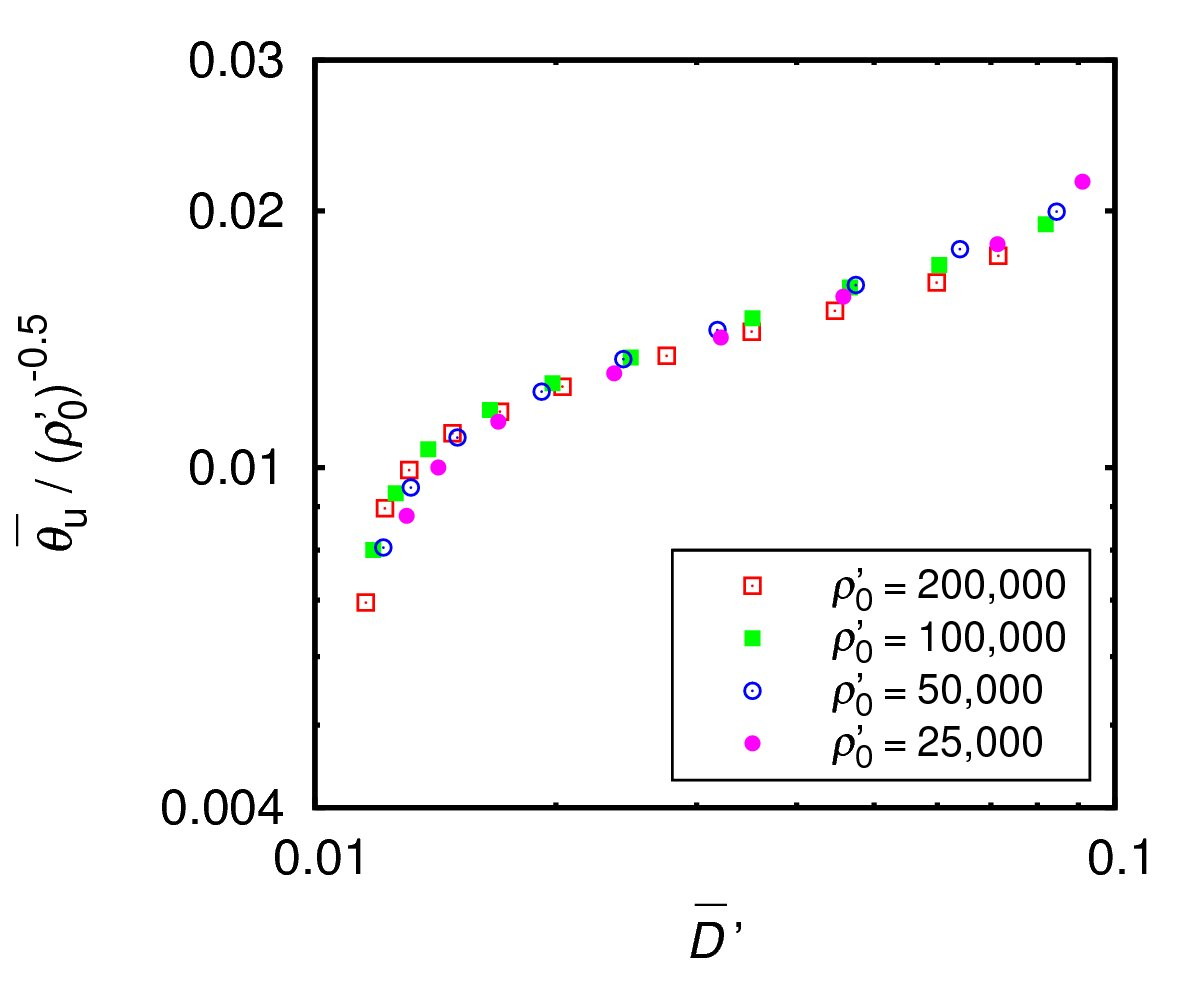}
\caption{\label{fig:davg_misorient_N} The average misorientation weighted by the initial average dislocation spacing as a function of the subgrain size. The lower is the initial density, the lower is the observed misorientation.}
\end{center}
\end{figure}

The contradiction can be removed if the initial dislocation density rather than the instantaneous dislocation density is considered in the scaling relation (\ref{eqn:holt}). In figure~\ref{fig:davg_misorient_N} the average misorientation $\overline{\theta}_\mathrm{u}$ was plotted as a function of the average subgrain size $\overline{D}$ at different initial dislocation densities $\rho_0$. The latter simply indicates the different number of initial dislocations in the $L\times L$ simulation box. In addition, the misorientation was divided by the initial average dislocation spacing. The curves fully overlap revealing $\overline{\theta}_\mathrm{u} = K \rho_0^{-0.5}$, with $K$ depending on time, average subgrain size, etc.

It is reasonable to assume that during the initial stages of the simulations, when the first dislocation walls are formed from the random arrangement of dislocations, the average distance of the dislocations within the walls must scale with $\rho_0^{-0.5}$. According to this discussion, this distance, which is directly related to $\overline{\theta}_\mathrm{u}$ through Frank's formula, remains nearly constant throughout the growth process. The slight increase observed can be related to the fact that the mobility of LAGBs is proportional to the misorientation \cite{humphreys}, therefore, LAGBs with higher misorientation wipe out the sparse ones (having lower misorientation). The picture which emerges is in complete agreement with the experiments, where the constancy of the misorientation is valid even for specimens with quite different initial $\overline{\theta}_\mathrm{u}$ \cite{huang_humphreys}.

Note that contrary to the present investigations, in the previous work of Bakó et al.~\cite{bako_2007}, Holt's relation was found to be satisfied throughout the growth process. After the more precise analysis of this paper, the conclusion obtained is that those results were due to numerical noise. The latter is similar to thermal noise, such that those results should be regarded as if very high temperature was applied (see figure~\ref{fig:davg_temp}).

The next question to be addressed is why Holt's relation is valid in a wide range of systems with cell structure. The possible reason is that in those cases subgrain formation is induced by external stress, rather than being a simple growth phenomenon. It is well-known, that during creep, the subgrain structure is characterized by a dynamic equilibrium, that is, the plastic strain is produced by the moving subgrain walls, but still, the average subgrain size remains constant \cite{kassner}. In this state, the net dislocation annihilation must be balanced by creation mechanisms. The latter is absent in this model and also during recovery. In summation, from this investigation it is evident that dislocation creation is the key to the dynamic equilibrium during creep, and to the fulfilment of Holt's relation.

\section{Conclusions}

Cellular dislocation patterning and its growth phenomenon has been investigated within a simple 2D discrete dislocation dynamics model. The effects of the climb rate, the thermal noise and the size of the dislocation core radius on the kinetics of growth have been studied. The main results of the paper are as follows:
\begin{enumerate}
\item The coarsening follows power-law kinetics, with a growth exponent not depending on the dislocation climb mobility but rather on the thermal noise and core radius.
\item The size distribution of the subgrains was close to lognormal, which broadened with time. This means that growth is abnormal (discontinuous).
\item Holt's relation is not fulfilled in the conventional way -- the average subgrain size is proportional to the initial dislocation spacing not to the temporal spacing.
\item The average misorientation between the adjacent subgrains slightly increases throughout the growth process.
\item Despite the simplicity of this model, all of the above results are in qualitative agreement with the experiments.
\item In contrast to previous dislocation climb-based models, our simulations yield growth exponents considerably larger than $n=2$.
\end{enumerate}
These results indicate the success of using 2D discrete dislocation dynamics in modelling subgrain coarsening.

\section*{Acknowledgement}

The financial support of the Hungarian Scientific Research Fund (OTKA) under Contract No.\ K~67778 is gratefully acknowledged.

\section*{References}

\bibliographystyle{unsrt}
\bibliography{./journalss.bib,./coarsening.bib}
%%\bibliography{journalss.bib,coarsening.bib}

\end{document}